\documentclass[aps,prb,twocolumn]{revtex4}

\usepackage{graphicx}
\usepackage{dcolumn}
\usepackage{amsmath}

\begin{document}

\title{Interplay between magnetic and lattice excitations in BiFeO$_3$}
\author{M. Cazayous}
\email{maximilien.cazayous@univ-paris-diderot.fr}
\author{A. Sacuto}
\affiliation{Laboratoire Mat\'eriaux et Ph\'enom\`enes Quantiques (UMR 7162 CNRS), 
Universit\'e Paris Diderot-Paris 7, 75205 Paris cedex 13, France}
\author{D. Lebeugle}
\author{D. Colson}
\affiliation{Service de Physique de l'Etat Condens\'e, DSM/DRECAM/SPEC, CEA Saclay, 91191 Gif-sur-Yvette, France}

\date{\today}
     
     
\begin{abstract}
We have performed Raman scattering investigations on the high energy magnetic excitations in a BiFeO$_3$ single crystal as a function of both temperature and laser excitation energy. A strong feature observed at 1250~cm$^{-1}$ in the Raman spectra has been previously assigned to two phonon overtone. We show here that its unusual frequency shift with the excitation energy and its asymmetric temperature dependent Fano lineshape reveal a strong coupling to magnetic excitations. In the same energy range, we have also identified the two-magnon excitation with a temperature dependence very similar to $\alpha$-Fe$_2$O$_3$ hematite.  
\end{abstract}

\maketitle
Recently, multiferroics have attracted much attention because they exhibit a strong coupling between 
magnetic and ferroelectric orders.\cite{Eerenstein} They are good candidates for controlling the 
magnetic properties by electric field and vice versa \cite{Zhao} making these materials 
promising for potential applications in spintronics and information storage.
BiFeO$_3$ (BFO) is one of the most studied multiferroic due to both its strong ferroelectric polarization \cite{Wang, Lebeugle, Lebeugle1} and antiferromagnetic order at room temperature\cite{Smolenski,Ismilzade,Sosnowska}. The origin of the magnetoelectric effect in BFO remains an open issue. In order to elucidate this question, investigating the interplay between lattice vibrations and magnetic excitations appears very useful because lattice distortions strongly affect the ferroelectric polarization and therefore its coupling to magnetic order. 
\par
THz spectroscopies have recently shown to be very efficient tools for probing the coupling between the lattice and magnetic excitations.
Raman\cite{Singh1, Fukumura, Cazayous} and infrared 
measurements\cite{Kamba, Lobo} combined with theoretical approaches\cite{Hermet} have permited to identify most of the vibrationnal normal modes in BFO.  At the same time 
low-frequency magneto-optical resonances in the dielectric susceptibility of multiferroic compounds have been observed
\cite{Pimenov, Mostovoy, Sushkov, Katsura} and recently the cycloidal magnetic modes have been optically detected in BFO\cite{Cazayous1, Singh}.
However, probing magnetic excitations and in particular their coupling to the lattice is still a challenge in BFO because we are still lacking experimental evidences for the entanglement with magnetic excitation through the phonon behaviors. 
\par
Here, we present Raman measurements on BFO single crystals aiming to address the interplay between lattice and magnetic orders. Among the two-phonon overtone features previously detected at high energy (above 700 cm-1), we show that the mode located at 1250~cm$^{-1}$  is much more complex than a simple double vibrational mode and exhibits a strong magnetic character. Indeed we reveal that the 1250~cm$^{-1}$ mode depends on the laser excitation energy and exhibits an asymmetric lineshape controlled by the spin dynamic of BFO. We have also detected in the same energy range, the signature of a two magnon mode near 1050~cm$^{-1}$ , its temperature-dependence is in agreement with the one observed in $\alpha$-Fe$_2$O$_3$ hematite .

\par
BiFeO$_3$ single crystals were grown in air using a Bi$_2$O$_3$-Fe$_2$O$_3$ 
flux technique.\cite{Lebeugle} They are typically of millimeter size.  
Below a high Curie temperature ($T_c$$\sim$1100 K) BFO has a rhombohedrally distorted perovskite 
structure with a spontaneous electrical polarization (between 50 and 100 $\mu$C/cm$^2$ in magnitude)   along the [111] direction in the pseudo-cubic configuration.\cite{Lebeugle} 
Below the N\'eel temperature ($T_N$$\sim$640 K)\cite{Smolenski,Ismilzade}, BFO exhibits a G-type antiferromagnetic order 
subjected to long range modulation associated with a cycloidal spiral with a 
period length of 62~nm\cite{Sosnowska}. The spiral propagation direction is along [10-1] with a spin rotation within (-12-1) plane.
\par
Raman measurements have been performed using 
several laser excitation lines between 1,92 to 2.41 eV from a mixed Ar$^+$-Kr$^+$ gas laser. 
The BFO crystals were mounted in vacuum (10$^{-6}$mbar) on the cold finger of a liquid helium
cryostat. The Raman spectra were recorded between 7 and 300~K using 
a triple grating spectrometer (JY-T64000) equipped with a nitrogen cooled CCD detector.
The spectrometer was in substractive configuration with a resolution of 0.5 cm$^{-1}$.
The Raman spectra have been obtained using unpolarized excitation light. 
\par

\begin{figure}
\includegraphics*[width=7cm]{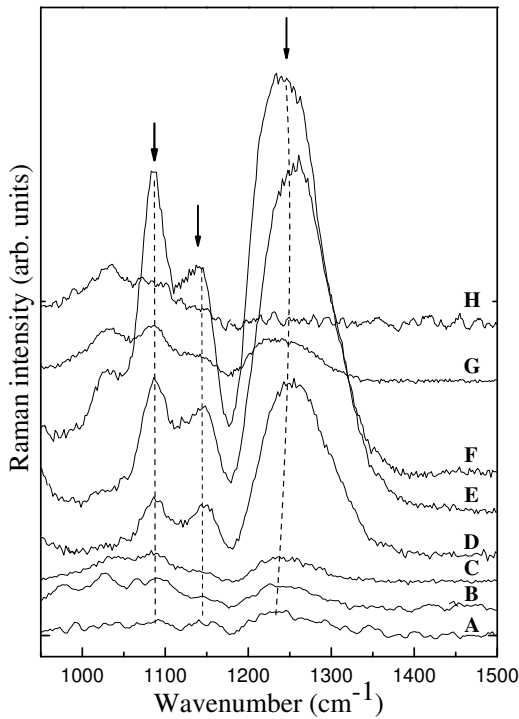}\\
\caption{\label{Figure1} Raman spectra of BiFeO$_3$  single crystal in the 900-1500~cm$^{-1}$ 
frequency range at 20K. The spectra labeled A to H were obtained from 2.71, 2.65, 2.6, 2.54, 2.41, 2.34, 2.18, 1.92~eV excitation laser lines, respectively. The Raman peaks of are indicated by arrows. Dashed lines are guides for the eyes. All spectra have been corrected for the spectral response of the spectrometer.}
\end{figure}

\begin{figure}
\includegraphics*[width=7cm]{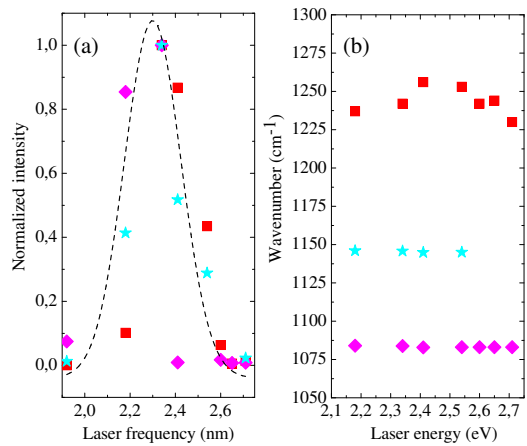}\\
\caption{\label{Figure2} (a) Intensities normalized to its maximum value and (b) wavenumber shifts as a function of the laser energy for the
1085 (diamond), 1140 (star) and 1250~cm$^{-1}$ (square) peaks.}
\end{figure}

\begin{figure}
\includegraphics*[width=7cm]{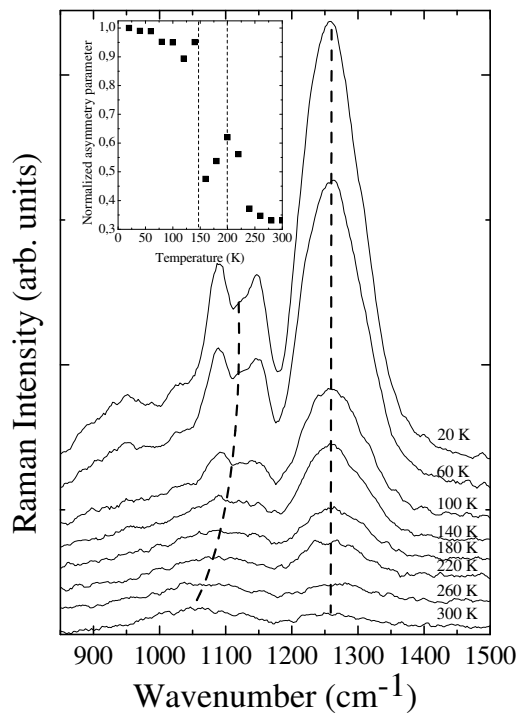}\\
\caption{\label{Figure3} Temperature dependance of the two-magnon and 1250~cm$^{-1}$ peaks obtained with the 2.34~eV laser line. 
Dashed lines are guides for the eyes. The peak at 1250~cm$^{-1}$ has been fitted by the Fano equation (see text).
The inset shows the temperature evolution of the asymmetric parameter $q$ corresponding to the function normalized 
to its maximum value.}
\end{figure}

\begin{figure}
\includegraphics*[width=8cm]{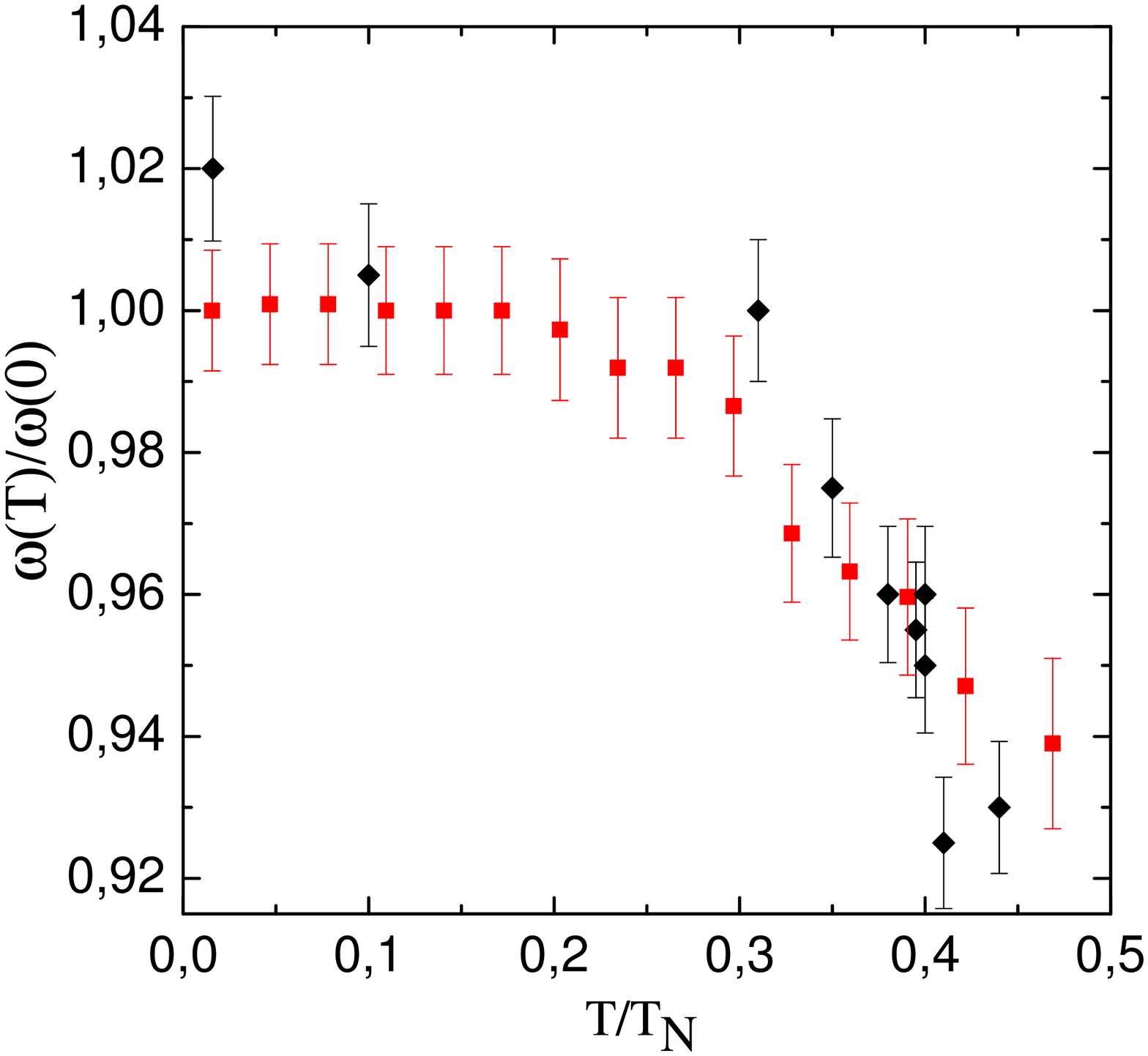}\\
\caption{\label{Figure4} \textbf{Temperature dependence} of the normalized two-magnon frequencies for BiFeO$_3$ (square) 
and hematite $\alpha$-Fe$_2$O$_3$ (diamond).}
\end{figure}

The assignment of the vibrationnal modes under 700~cm$^{-1}$ is well 
known.\cite{Singh1, Cazayous, Fukumura} Group theory predicts 13 Raman active phonons 
for the R3C symmetry of BFO: $\Gamma=4A_1+9E$. Above 700~cm$^{-1}$, one should only detect the two-phonon Raman peaks.
Two-phonon Raman scattering is a second order process involving two phonons.
In contrast to one-phonon scattering process there is no restriction on the phonon wavevector
 but the sum of the two phonon wavectors has to be zero.  If the two phonons are identical, the two-phonon Raman peak is called an overtone. 
\par
Figure~\ref{Figure1} shows the Raman spectra in the 1000-1300~cm$^{-1}$ range as a function of the laser excitation energy. 
We focus here on the three main peaks at 1085, 1140 and 1250~cm$^{-1}$ (indicated by arrows in Fig.~\ref{Figure1}). 
The Raman peak intensities are strongly enhanced under the blue (2.34~eV) excitation line. 
These enhancements or resonance come from the matching of the incident photon energy with the absorption edge of BFO.\cite{Fukumura, Ramirez} 
As the incident photon energy is close to the absorption edge, the probability of creating electron-hole pairs diverges and the Raman scattering efficiency is greatly enhanced. 
\par
The three peaks have been assigned to a two phonon Raman scattering process.\cite{Ramirez} 
In particular, the prominent peak at 1250~cm$^{-1}$ has been previously reported by 
Fukumura {\it et al.}. Its sudden disappearence has been observed around 600~K-700~K close to $T_N$
pointing out it's probably magnetic origin.\cite{Fukumura1}  
More recently, Ramirez {\it et al.} have assigned this peak to a pure two phonon process (2x[E$_{9}$(TO)$\cong$ 620~cm$^{-1}$]).\cite{Ramirez} 
However, the decrease of its integrated intensity up to $T_N$ and its constant behaviour beyond $T_N$ advocate in favor of a strong coupling between the ferroelectric and magnetic subsystems. 
\par
In Fig.~\ref{Figure1}, the two phonon modes locations at 1085 and 1140~cm$^{-1}$ remains constant in frequency as the excitation energy is tuned.  
On the contrary, the frequency of the 1250~cm$^{-1}$ peak increases when the excitation energy is tuned from 2.71 to 2.41~eV and 
then decreases. The frequency shift is about 20~cm$^{-1}$ as shown in Fig.~\ref{Figure2}(b).
To explain the different behaviour with the excitation energy of the 1250~cm$^{-1}$ overtone compare to the 1085 and 1140~cm$^{-1}$ one
 different scenarii can be put forward:
\par
(i) The peak at 1250~cm$^{-1}$ is broad and might be constituted of two peaks with two different resonant behaviors with the excitation energy thus explaining the observed shift of the 1250~cm$^{-1}$ peak in Fig.~\ref{Figure1}. 
The frequency of these two peaks located at 1228 and 1288~cm$^{-1}$ can be deduced from a 
gaussian fit deconvolution of the broad 1250~cm$^{-1}$ peak. The 1228 and 1288~cm$^{-1}$ peaks
should result respectively from an overtone of a peak around 615 and 645~cm$^{-1}$.  
However, only one peak has been detected around 620~cm$^{-1}$ and we are led to conclude that the peak observed at 1250~cm$^{-1}$ is unlikely composed of two peaks.
\par
(ii) The Raman band is associated with a luminescence process assisted by magnons or phonons.
Strong exciton-phonon interactions lead to new exciton-phonon bound states which
causes phonon-assisted exciton transitions. For luminescence, the transitions via the new bound states 
induce a continuous background which interferes with the usual discrete transitions imply in the luminescence process. 
This would be the origin of Fano lineshape in the phonon-assisted luminescence process.
Since luminescence is produced by thermalized electron-hole pairs, 
the emmitted photons have no correlation with the excitation lines. 
As a result, the frequency of the luminescence peak is independant of the excitation energy 
and increases monotonically with the incident photon energy. 
The frequency maximum of the 1250~cm$^{-1}$ peak in Fig.~\ref{Figure2}(b) is not consistent with this interpretation.
\par
(iii) Such a behaviour is due to the coupling of the peak at 1250~cm$^{-1}$ with the antiferromagnetic order
as suggested by Fukumura {\it et al.} and Ramirez {\it et al.}.
 \par
In order to analyse and identify the different contributions of the three main peaks detected in the 1000-1300~cm$^{-1}$ range of the Raman spectra, we have preformed Raman measurements as a function of temperature (from 20 to 300~K) as shown in  Fig.~\ref{Figure3}. 
The three main peaks exhibit a small softening in energy (about 5~cm$^{-1}$), gradually broaden and decrease 
in intensity as the temperature is raised. The peak at 1250~cm$^{-1}$ presents an asymmetric temperature dependent lineshape characteristic of a strong coupling between the two phonon mode and the spin excitation spectrum as we will see just below. 
\par
In order to quantify the Fano lineshape we have fitted the Raman peak located at 1250~cm$^{-1}$  by the Fano equation:\cite{Fano}
$I(\omega) = I_0\frac{(q+\frac{(\omega-\omega_0)}{\Gamma})^2}{1+\frac{(\omega-\omega_0)}{\Gamma})^2}$ 
where $\omega_0$ is the peak position without interaction, $q$ is the Fano asymmetric parameter and $\Gamma$ is the  width at half maximum. 
The $q$ parameter serves as a measure of the coupling between the two phonon state and a continuum of states that we have to identify.

\par
The evolution of the normalized asymmetric parameter $q$ as a function of the temperature is plotted in the inset of Fig.~\ref{Figure3}.
A first glance on the temperature evolution of the $q$  parameter reveals two distinct regimes: $q(T)$ 
exhibits a sharp step close to $T=150~K$.
\par
Recent measurements on spin waves in BFO have shown a spin-reorientation close to these two temperatures.\cite{Cazayous1, Singh} 
The $q$ parameter is thus sensitive to the spin-reorientation which points out the intimate relationship between the double phonon peak at 1250~cm$^{-1}$ and the spin sublattice excitations. 
From 10~K to 150~K the asymmetric parameter $q$ remains constant and keeps its high value indicating that the two-phonon mode is strongly coupled to the spin excitations in this temperature range. 
Around 150~K, the spin-phonon coupling is modified by the reorientation of the spins and sharply decreases as shown by the decrease of the $q$ parameter.  
An additional feature acures at 200~K where another spin-reorientation has been observed.\cite{Singh}
Above 200~K the asymmetry parameter decreases continuously as a mirror of the 
softning of the spin-phonon coupling due to thermal excitations.  
\par
In Fig.~\ref{Figure3}, the two phonons peaks at 1085 and 1140~cm$^{-1}$ disappear at 140~K 
leaving a broad and weak peak. From 140~K to 300~K this broad peak shifts from 1110~cm$^{-1}$ down to 
1050~cm$^{-1}$. The large energy softening (about 60~cm$^{-1}$) of the peak with temperature 
is unusual compared to the one of the phonons (about 5~cm$^{-1}$ as  mentionned before). 
We suspect that this large feature is related to a magnetic excitation and thus sensitive to T$_N$. 
\par
In order to establish the origin of this broad peak, we have studied the temperature dependence of its energy. 
In Fig.~\ref{Figure4} is plotted  the normalized frequency shift versus the temperature normalized to T$_N$.
The broad peak contibution in Fig.~\ref{Figure3} has been obtained substracting the gaussian fit contribution of the phonon peaks.  
 We observe  softening of the freqeuncy of the presumed two-magnon peak as a function of temperature quite consistent with the temperature dependance of the two-magnon peak previously measured in hematite $\alpha$-Fe$_2$O$_3$, FeBO$_3$ and MnF$_2$\cite{Massey}. 
For clarity, only the behaviour of the two-magnon peak in hematite has been plotted in Fig.~\ref{Figure4}.
The good agreement between the two curves strongly supports the assignment of the broad
 peak centered at 1050~cm$^{-1}$ (300~K) to a two-magnon peak.
\par
The two-magnon scattering involves two-magnons with  equal 
and opposite wave vectors due to the momentum conservation similar to  the two-phonon Raman scattering process .\cite{Fleury}  
The peak position related to the two-magnon scattering is expected equal to twice the zone boundary magnon energy and can be
compared to the one measured in orthoferrites (RFeO$_3$ with R a rare-earth atom).
In YFeO$_3$ and ErFeO$_3$ the two-magnon band is measured around 900~cm$^{-1}$.\cite{Koshizuka,Takahashi}
\par
We point out that the two-magnon signature should also be observed in infrared spectroscopy. 
In some crystal with a center of inversion the two-magnon excitation through the exchange dipole 
moment is not allowed but can be observed using the oscillation strength of the phonon. In BFO, no inversion symmetry is found between the Fe$^{3+}$ ions which allow them to make significant contributions to the direct optical excitation of the two magnons. 
\par
In summary, we have tracked the magnetic excitations of the high energy Raman spectra. The complex nature of the 
1250~cm$^{-1}$ feature have been revealed. This peak is an overtone of two phonon band with a frequency and a temperature dependance  
driven by its interaction with the magnetic subsystem. In addition, we have measured a weak and broad peak close to 1050~cm$^{-1}$ 
whose temperature dependence is consistent with two-magnon scattering process. \\

\end{document}